\documentclass[aps,prl,twocolumn,groupedaddress,showpacs]{revtex4}

\usepackage{graphicx}
\newcommand{\myw}{0.76}

\begin{document}

\author{S. E. Pollack}
\author{D. Dries}
\author{M. Junker\footnote{Current Address: School of Physics, University of Melbourne, Victoria 3010, Australia}}
\author{Y. P. Chen\footnote{Current Address: Dept of Physics, Purdue University, 525 Northwestern Ave., West Lafayette, IN 47907}}
\author{T. A. Corcovilos}
\author{R. G. Hulet}
\affiliation{Department of Physics and Astronomy and Rice Quantum Institute, 
    Rice University, Houston, TX  77005}

\date{\today}

\title{Extreme tunability of interactions in a $^7$Li Bose-Einstein condensate}

\begin{abstract}
We use a Feshbach resonance to tune the scattering length $a$ of 
a Bose-Einstein condensate of $^7$Li in the $|F = 1, m_F = 1\rangle$ state.
Using the spatial extent of the trapped condensate
we extract $a$ over a range spanning 7 decades from
small attractive interactions to extremely strong repulsive interactions.
The shallow zero-crossing in the wing of the Feshbach resonance
enables the determination of $a$ as small as $0.01$ Bohr radii.
Evidence of the weak 
anisotropic magnetic dipole interaction 
is obtained
by comparison with different trap geometries for small $a$.
\end{abstract}


\maketitle


The ability to control the parameters
of ultracold atomic gases 
and to impose external potentials upon them
provides unique
opportunities to create model systems for exploring complex phenomena in 
condensed matter and nuclear physics.  
Control of atomic interactions using Feshbach resonances has proven to be 
particularly productive in studies involving Bose-Einstein condensates (BECs) 
or paired Fermi gases \cite{frucg1}.
While strong interactions are usually the focus of these studies, 
interesting phenomena also occur in the weakly-interacting regime.  
An example of such a phenomenon is 
Anderson localization in disordered media \cite{PhysRev.109.1492},
which was recently observed in weakly repulsive BECs \cite{AspectAL, ModugnoAL}.
Another example is the formation of bright solitons in BECs
with weakly attractive interactions, 
which have been created in condensates of 
$^7$Li \cite{Khaykovich05172002, StreckerSolitons}
and $^{85}$Rb \cite{cornish:170401}.
Atom interferometers may also benefit by the increased coherence times afforded by weakly
interacting gases \cite{gustavsson:080404, fattori:080405},
or even by a dispersionless atomic soliton laser \cite{StreckerSolitons, PhysRevA.70.033607}.

Several atomic species exhibit Feshbach resonances where the 
$s$-wave scattering length $a$ changes sign at a certain field in the wings of the resonance.  
These zero-crossings are useful in applications requiring weak interactions.
In addition to $^7$Li \cite{Khaykovich05172002, StreckerSolitons},
such zero-crossings have been studied in 
$^{85}$Rb \cite{PhysRevLett.85.1795}, 
$^{52}$Cr \cite{Lahaye},
$^{39}$K \cite{roati:010403}, and
$^{133}$Cs \cite{gustavsson:080404}, 
In this Letter, we report the measurement of $a$ for 
$^7$Li in the $|F = 1, m_F = 1\rangle$ state for fields near the Feshbach 
resonance at 737\,G \cite{Khaykovich05172002, StreckerSolitons, PhysRevA.51.4852, junker}.
By measuring the \textit{in situ} size of the confined condensate, 
$a$ is measured over a range of 7 decades.  
We find that the slope of the zero-crossing is only 
$\sim$0.1\,$a_0$/G, where $a_0$ is the Bohr radius.  
This is the shallowest known zero-crossing, requiring only modest field 
stability to achieve an essentially non-interacting gas.  
We explore the effects of the magnetic dipole interaction (MDI)
in this regime.
Unlike Cr, which has a large magnetic moment of $6\,\mu_\mathrm{B}$
resulting in a relatively large MDI \cite{Cr, koch:218, lahaye:080401},
the MDI in alkali atoms is weak due to their small
magnetic moments of $\sim$$1\,\mu_\mathrm{B}$.
Nonetheless, the MDI has been recently detected in alkali atoms \cite{fattori:190405, vengalattore:170403}.
We explore the role of the MDI by modifying the
confining geometry of the BEC.


Our experimental apparatus for generating a BEC of $^7$Li
has been described previously \cite{StreckerSolitons, Disorder}.
Atoms in the $|F=1, m_F=1\rangle$ state are confined in an optical trap
formed from a single focused laser beam with wavelength of 1.03\,$\mu$m.
A bias magnetic field, directed along the trap axis, is used to tune $a$ via the Feshbach resonance.
We create condensates at a field where $a$ is large to facilitate rapid rethermalization
of the atoms during evaporation from the optical dipole trap.
After a condensate is formed we slowly ($\sim$4\,s) 
ramp the field to the desired value and determine the
scattering length, as described below.
There is no discernable thermal part to the density distributions
and we estimate that $T/T_c < 0.5$, where $T_c$ is the condensation temperature.
The final trapping potential is a combination of the optical field and a
residual axial magnetic curvature from the bias field.
The trap is cylindrically symmetric with measured radial and axial trapping frequencies of 
$\omega_r/2\pi = 193\,$Hz and $\omega_z/2\pi = 3\,$Hz, respectively.

\begin{figure}
  \includegraphics[width=1.0\columnwidth,angle=0]{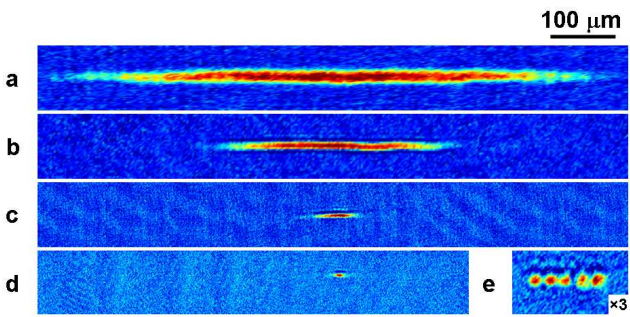}
\caption{
(color online) 
Representative \textit{in situ} polarization phase-contrast images of
condensates with various interaction strengths.
(a) $B=719.1\,$G, $a = 396\,a_0$, $N=1.7 \times 10^5$;
(b) $B=597.4\,$G, $a = 8\,a_0$, $N=2.9 \times 10^5$;
(c) $B=544.7\,$G, $a = 0.1\,a_0$, $N=2.0 \times 10^5$;
(d) $B=542.4\,$G, $a = -0.1\,a_0$, $N=1.2 \times 10^5$;
(e) same as (d) but with a faster field ramp from 
$710\,$G to $542.4\,$G, resulting in multiple solitons with $N\approx10^4$ per soliton.
The probe laser detuning from resonance is adjusted to keep a nearly constant signal level,
and varies between $20\,\gamma$ for large $a$ to $150\,\gamma$ for small $a$,
where $\gamma/2\pi \approx 5.9\,$MHz is the excited state linewidth. 
The color map is adjusted to maximize contrast for each image.
\label{fig:images}}
\end{figure}

We use \textit{in situ} polarization phase-contrast imaging 
\cite{PhysRevLett.78.985} to acquire the column density distribution 
of the condensate at the desired magnetic field.
When the $s$-wave interactions are large and repulsive they inflate the size of
the condensate well above the harmonic oscillator size.  
As the interactions decrease the size of the condensate becomes
smaller, approaching the harmonic oscillator ground state near zero interactions.
Figure~\ref{fig:images} shows representative images of condensates
with various repulsive or attractive interaction strengths.
Solitons form when $a < 0$, either a single one for a slow magnetic
field ramp or multiple solitons for ramps fast compared to the axial trap period.

\begin{figure}
  \includegraphics[width=\myw\columnwidth,angle=-90]{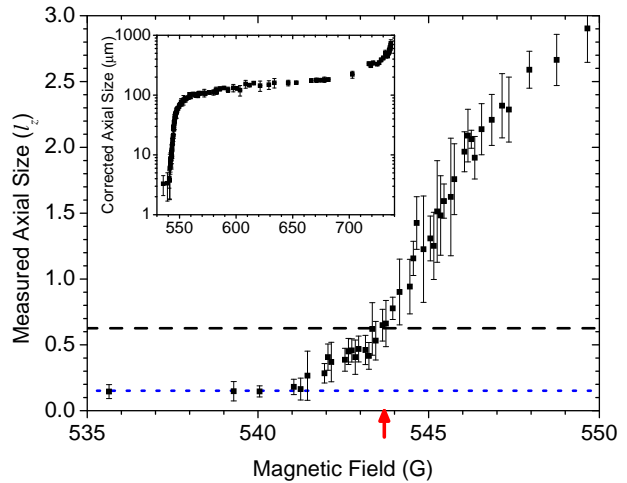}
\caption{(color online)
Axial size of the condensate as a function of magnetic field.
The axial size is defined as the $1/e$ radius of the axial density profile
and is scaled by the axial harmonic oscillator size 
$l_z = \sqrt{\hbar / m \omega_z} \approx 22\,\mu$m.
The resolution of the optical imaging system is $\sim$3.3$\,\mathrm{\mu m}$ (dotted line).
The dashed line is the size of the condensate ($l_d\approx0.62\,l_z$)
found by solving Eq.~(\ref{eqn:gpe}) with $a = 0$. 
The zero-crossing in $a$ occurs when the size of the condensate equals $l_d$ (arrow).
Neglecting dipolar effects results in a zero-crossing about 0.5\,G higher, where
the axial size equals $l_z$.
The inset shows the axial size corrected for number variation as described in the text.
Individual data points and error bars are the 
average and standard error of approximately 10 shots taken at each field.
Systematic uncertainty in the axial size is $\sim$3\% from 
uncertainty in temperature and 
the uncertainty in imaging magnification.
The systematic uncertainty in magnetic field due to calibration
(via radio frequency transitions from the $|2,2\rangle$ to the $|1,1\rangle$ state)
is $\sim$0.1\,G.  We have binned the data into intervals of this size.
\label{fig:axialSize}}
\end{figure}

We integrate the image of the 
condensate in the remaining radial dimension to produce
an axial density profile. 
The $1/e$ radius of this profile is used as a measure of the condensate size,
as shown in Fig.~\ref{fig:axialSize}
for a range of magnetic field values.  
In the Thomas-Fermi regime, the axial size of the condensate is dependent on the
product of $a$ and the number of atoms in the condensate $N$.
The average number per condensate is $N_0 = 3 \times 10^5$ atoms, 
with a shot-to-shot variation of 20\%.
The inset of Fig.~\ref{fig:axialSize} shows
the axial size scaled by $(N/N_0)^{1/5}$ to account for these fluctuations.  
Several condensates are found to have axial sizes smaller than 
the axial harmonic oscillator size 
due to net attractive interactions, as discussed below.

\begin{figure}
  \includegraphics[width=\myw\columnwidth,angle=-90]{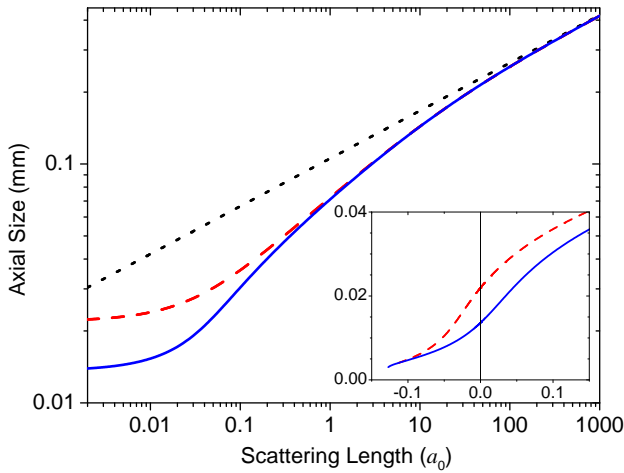}
\caption{(color online)
Mapping functions of axial size to $a$
using a Gaussian trial wavefunction in a
variational solution to Eq.~(\ref{eqn:gpe}),
including (solid) and neglecting (dashed) the MDI;
also shown is the Thomas-Fermi approximation (dotted).
These mappings were computed for $N = 3 \times 10^5$, $\omega_r/2\pi = 193\,$Hz
and $\omega_z/2\pi = 3\,$Hz.  In practice,
we compute the mapping individually for each imaged condensate to 
account for variations in $N$ and a field dependent variation in $\omega_z$
of $\sim$5\% over the relevant magnetic field range.
The Gaussian solution neglecting the MDI asymptotically
approaches $l_z$ at zero interactions,
while their inclusion causes the solution to 
asymptotically approach a value smaller than $l_z$.
\label{fig:mapping}}
\end{figure}

To determine $a$ for each image
requires a mapping from the measured axial size and $N$ to $a$.  
We model the system using the three-dimensional (3D) Gross-Pitaevskii equation
\begin{eqnarray}\label{eqn:gpe}
\mu \psi =&& -\frac{\hbar^2}{2 m}\nabla^2 \psi + V \psi + \frac{4 \pi \hbar^2 a}{m} |\psi|^2 \psi \nonumber\\
&& + \frac{\mu_0 \mu_m^2}{4 \pi} \int \frac{1 - 3 \cos^2 \theta}{|\mathbf{r}-\mathbf{r}'|^3} |\psi (\mathbf{r}')|^2 d\mathbf{r}'\psi,
\end{eqnarray}
where we account for the MDI
in addition to the $s$-wave contact interaction and the trapping potential.
At 540\,G, $\mu_m \approx 0.94\,\mu_\mathrm{B}$
for $^7$Li in the $|1,1\rangle$ state.
Mapping is accomplished by performing a variational calculation
using a 3D cylindrically symmetric Gaussian wavefunction as a trial solution to Eq.~(\ref{eqn:gpe}).
Minimizing the corresponding energy functional results in equations for 
the radial and axial sizes of the condensate \cite{PhysRevA.63.053607},
which are solved to give the desired mapping function.  Figure~\ref{fig:mapping}
shows this mapping with and without inclusion of the MDI,
as well as the corresponding mapping from the Thomas-Fermi approximation.
In our geometry, the magnetic moments are aligned with the long axis of the trap.
This causes the MDI to be effectively attractive, 
making the condensate smaller axially for a given value of $a$.  
We have verified the accuracy of the variational calculation by
exact numerical solution of
Eq.~(\ref{eqn:gpe}) for various values of $a$ and 
find good agreement between the two methods.
Since the variational calculation is much faster computationally, 
we use this method to analyze the data.

\begin{figure}
  \includegraphics[width=.8\columnwidth,angle=-90]{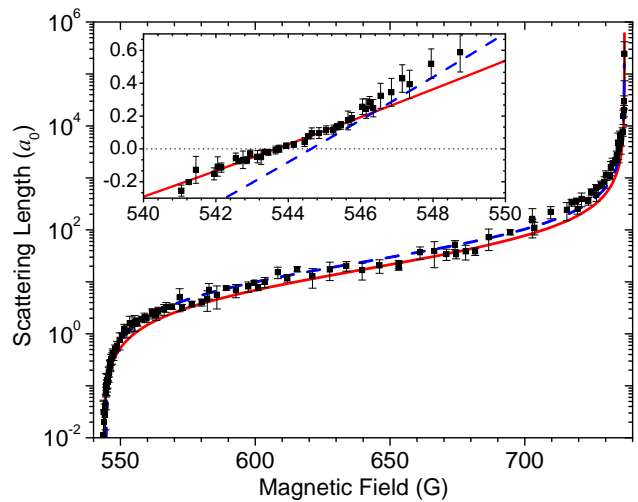}
\caption{(color online)
Axial size data of Fig.~\ref{fig:axialSize} mapped onto $a$.
Results of a coupled-channels calculation are shown by the solid line.
The Feshbach resonance fit is indicated by the dashed line.
The inset shows the extracted values of $a$ near the zero-crossing.  
The mean and standard error of approximately 10 shots
taken at each field is shown.  
In addition, we estimate a systematic uncertainty of $\sim$20\% in $a$.
\label{fig:ascatt}}
\end{figure}

Figure~\ref{fig:ascatt} shows the axial size data of Fig.~\ref{fig:axialSize} mapped onto $a$.
The general shape follows that of a typical Feshbach resonance with
$a = a_{BG} [1 + \Delta / (B - B_{\infty})]$, where
$a_{BG} = -24.5^{+3.0}_{-0.2}\,a_0$,
$\Delta = 192.3(3)\,\mathrm{G}$,
and $B_{\infty} = 736.8(2)\,\mathrm{G}$.
The uncertainties in these derived values are a result of the
systematic uncertainty in field calibration of 0.1\,G
and a systematic uncertainty in $a$ of $\sim$20\%, primarily due to uncertainty 
in measuring the axial size and determination of $\omega_z$.
A linear fit to the data for $B < 550\,$G gives
a slope of $0.08(1)\,a_0/\mathrm{G}$
and a zero-crossing at $B_0 = 543.6(1)$\,G \cite{error}.
The smallest mean positive scattering length of a collection of shots
was $0.01(2)\,a_0$ at 543.6(1)\,G with $\sim$$3\times 10^5$ atoms.
Under these conditions the peak density is $3 \times 10^{14}\,\mathrm{cm}^{-3}$
and the corresponding condensate healing length 
is comparable to the length of the condensate itself.
Although Eq.~(\ref{eqn:gpe}) assumes the mean field approximation,
beyond mean field corrections are expected to be important 
when $n a^3 \gtrsim 1$
\cite{HY, LY, LHY, PhysRevA.63.063601, papp:135301}.
The leading order correction to the interaction term in Eq.~(\ref{eqn:gpe}), 
the Lee-Huang-Yang parameter, 
is $\alpha = 32/(3\sqrt{\pi}) \sqrt{n a^3} \gg 1$
for the most strongly interacting condensates observed.
We have accounted for this correction in extracting $a$ for data where $\alpha < 1$.
For the four data points with $\alpha > 1$, this low-density expansion is not valid.
We are unaware of a theoretical treatment that addresses the 
density distribution in the strongly interacting regime.
While we extract a value a $a$ for these four data points 
by fitting to a Thomas-Fermi profile ignoring beyond mean-field effects, and plot them in Fig.
4, we exclude them in the Feshbach resonance fit.
Using this method
the largest mean positive scattering length was $\sim$$2 \times 10^5\,a_0$ 
at 736.9(1)\,G with $\sim$$2 \times 10^4$ atoms,
which has a peak density $n \approx 5 \times 10^{10}\,\mathrm{cm}^{-3}$.
The comparatively smaller number of atoms close to resonance is likely due to large
inelastic collisional losses in this regime \cite{PhysRevLett.77.2921}.
 
Figure~\ref{fig:ascatt} also shows a comparison between a coupled-channels calculation 
and the experimentally extracted values of $a$.  
The coupled-channels calculation requires
the ground-state singlet and triplet potentials of $^7$Li$_2$ as input, 
as described previously \cite{PhysRevLett.74.1315, PhysRevA.55.R3299}.  
We have updated the potentials to be consistent with the new measurements of
$B_\infty$ and $B_0$ reported here, 
as well as the previously measured binding energy of the 
least bound triplet vibrational level \cite{PhysRevLett.74.1315, footnote}.
The updates involve adjusting the singlet and triplet dissociation energies to
$D_e(X^1\Sigma^+_g) = 8516.68(10)\,\mathrm{cm}^{-1}$ and
$D_e(a^3\Sigma^+_u) = 333.714(40)\,\mathrm{cm}^{-1}$, 
where the stated uncertainties account for 
uncertainties in the remaining portions of the model potentials.
These values are consistent with previous determinations
\cite{PhysRevA.55.R3299, linton:6036, colavecchia:5484}.
The agreement between the calculated and measured values of $a$, 
while not perfect over the entire range of fields, 
is reasonably accurate over a range spanning 7 decades.


\begin{figure}
  \includegraphics[width=1.03\columnwidth,angle=0]{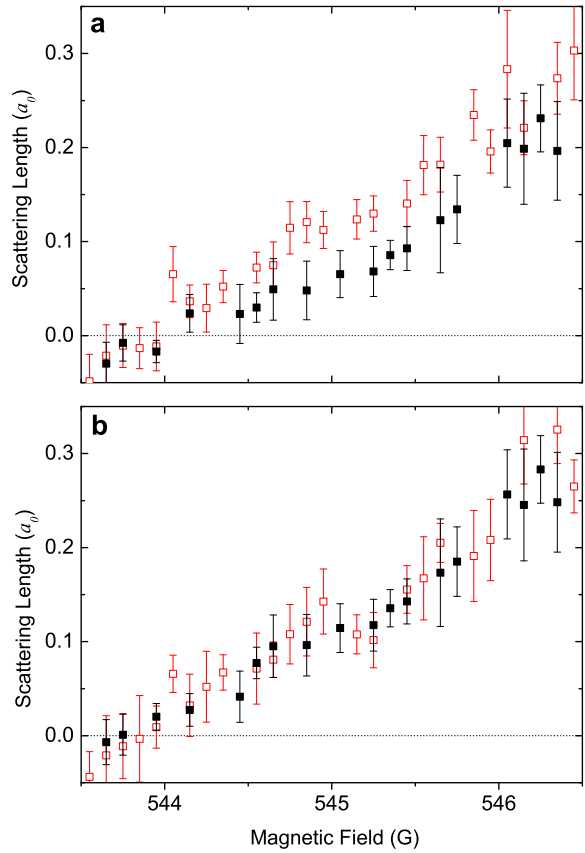}
\caption{
(color online) 
Extracted values of $a$ near the zero-crossing for trapping potentials with
$\omega_z/2\pi = 3$\,Hz (filled squares) or
$\omega_z/2\pi = 16$\,Hz (unfilled squares),
when (a) neglecting or (b) including the MDI in the mapping function.
The MDI has a negligible effect on the extracted values of $a$ for the 16\,Hz trap,
but neglecting the MDI in analysis of the 3\,Hz trap 
systematically lowers the mapped values of $a$,
especially for $a \lesssim 0.15\,a_0$.
\label{fig:dipole}}
\end{figure}
The effects of the MDI are strongly dependent on geometry.
To better distinguish their role,
we increased the axial trapping frequency from 3\,Hz to 16\,Hz
by applying magnetic curvature.
Figure~\ref{fig:dipole} compares the extracted values of $a$ for both trap geometries
when the MDI is included or neglected in the mapping function.
As expected, neglecting the MDI in the analysis systematically lowers
the extracted values of $a$.
This effect is most noticeable near the zero-crossing
where a systematic geometry-dependent discrepancy appears in the derived
values of $a$. 
Inclusion of the MDI in the analysis produces a consistent value of $a$ for a given magnetic field
regardless of the trapping potential.
The data show that the magnetic dipole interaction, although
weak, is discernible in $^7$Li despite having a magnetic moment of only $\sim$$1\,\mu_\mathrm{B}$.

We have mapped the Feshbach resonance from the regime of small attractive interactions
far from the resonance to extremely strong repulsive interactions very close to resonance.
The zero-crossing and resonance positions have been precisely located, 
enabling experimental access to a broad range of accurately known interactions.
Of particular interest will be explorations of atom and soliton transport 
through a disordered potential in the weakly interacting regime.


\begin{acknowledgments}
We thank James Hitchcock and Chris Welford
for their contributions to this project.
Support for this work was provided by the NSF, ONR, 
the Keck Foundation, and the Welch Foundation (C-1133).
\end{acknowledgments}

\bibliography{zerocrossing}

\end{document}